\documentclass[a4paper,11pt]{article}
\pdfoutput=1 

\usepackage{jcappub} 

\usepackage[T1]{fontenc} 
\usepackage{amsmath}
\usepackage{caption}
\usepackage{hyperref}
\usepackage[utf8]{inputenc}
\usepackage{amssymb}
\usepackage{wrapfig}
\usepackage{multirow}
\hypersetup{
    colorlinks=true,
    linkcolor=blue,   
    urlcolor=blue,
}

\title{\boldmath On the prior dependence of cosmological constraints on some dark matter interactions}

\author{James A.~D.~Diacoumis}
\author{and Yvonne Y.~Y.~Wong}

\affiliation{School of Physics, The University of New South Wales, Sydney NSW 2052, Australia}

\emailAdd{j.diacoumis@unsw.edu.au}
\emailAdd{yvonne.y.wong@unsw.edu.au}

\abstract{We explore the issue of prior dependence in the context of one-sided constraints on the dark matter--photon and dark matter--neutrino elastic scattering cross-sections derived from cosmic microwave background (CMB) anisotropies measurements. Testing in particular the linear flat,  Jeffreys, and logarithmic flat priors, we find that the former two yield upper limits on the cross-sections that are mutually consistent to within 20\%. In contrast, bounds derived under the assumption of the logarithmic flat prior are strongly sensitive to the choice of the lower prior boundary. Indeed, surveying the recent literature, we find that this pathology of the logarithmic prior has resulted in published constraints that are up to an order of magnitude artificially tighter than they should objectively be. Our revised ``objective'' constraints from the 2015 data of the Planck CMB mission on the present-day scattering cross-sections are 
$\sigma_{\rm DM-\gamma} < 1.72 \times 10^{-6} \, \sigma_{\rm T} \, (m_{\rm DM}/{\rm GeV})$ and $\sigma_{\rm DM-\gamma}  <  2.74 \times 10^{-15} \, \sigma_{\rm T} \, (m_{\rm DM}/{\rm GeV})$ for dark matter--photon interactions scaling as $a^0$ and $a^{-2}$ respectively, where $a$ is the scale factor, and $\sigma_{\rm T}$ the total Thomson scattering cross-section. Their dark matter--neutrino counterparts read 
$\sigma_{\rm DM-\nu} < 2.14 \times 10^{-6} \, \sigma_{\rm T} \, (m_{\rm DM}/{\rm GeV})$ and $\sigma_{\rm DM-\nu}< 2.46 \times 10^{-15} \, \sigma_{\rm T} \, (m_{\rm DM}/{\rm GeV})$. All have been computed assuming the  Jeffreys prior.
}

\begin{document}
\maketitle
\flushbottom
\renewcommand{\arraystretch}{1.4}

\section{Introduction}

Cosmology has, for many years, been a ``vanilla'' science in the sense that a large array of observations can be simultaneously well-described---to percent level precision---by a small set of model parameters. In the most basic analysis, the six variables that are tested against observational data are the baryon density \(\Omega_{b} h^{2}\), the cold dark matter density~\(\Omega_{c} h^{2}\), Hubble expansion rate~\(H_{0}\), optical depth to reionisation \(\tau_{\textrm{reio}}\), and the spectral index \(n_{s}\) and amplitude~\(A_s\) of the primordial curvature power spectrum. Coupled with the assumptions of spatial flatness and negligible primordial tensors, these variables and the values accorded to them by observations make up the so-called concordance flat \(\Lambda \textrm{CDM}\) model~\cite{Aghanim:2018eyx, Planck}.

Given its enormous success, it is natural to ask, to what limits can \(\Lambda \textrm{CDM}\) be pushed through the relaxation of model assumptions and/or the incorporation of new physical phenomena. Such inquiries are most commonly explored by means of small excursions around the base six-variable $\Lambda$CDM fit in the form of one-variable additions: in this way, one searches for an improved fit to the data in the presence of the additional degree of freedom that also preserves to a large extent the successful features of the base $\Lambda$CDM fit. Examples of such one-variable additions commonly found in the literature include neutrino mass, the primordial tensor-to-scalar ratio, and a non-canonical dark energy equation of state parameter.

We focus in this work on a specific one-variable extension to the base $\Lambda$CDM model, wherein the dark matter (DM) is endowed with elastic scattering with a standard model radiation component (photons or neutrinos). The relevant variable is $\sigma_{{\rm DM}-X}/m_{\rm DM}$, where $\sigma_{{\rm DM}-X}$, with $X = \gamma, \nu$, is the DM-$X$ elastic scattering cross-section, and $m_{\rm DM}$ is the dark matter mass. Such a scenario has been studied extensively in the context of solving the so-called small-scale crisis of cold dark matter~\cite{InteractingDM, InteractingDM2}, and upper bounds on $\sigma_{{\rm DM}-X}/m_{\rm DM}$ have been obtained and/or projected for a wide variety of cosmological observables. These include the cosmic microwave background (CMB) anisotropies~\cite{Wilkphoton,Wilknu,Stadler:2018jin}, Lyman-\(\alpha\) forest~\cite{Wilknu}, Milky Way satellite counts~\cite{Boehm:2014vja, Escudero:2018thh}, CMB spectral distortions~\cite{Ali-Haimoud:2015pwa, Diacoumis:2017hff}, $B$-mode~\cite{Ghosh:2017jdy}, and large-scale structure surveys~\cite{Escudero:2015yka,Diacoumis:2018nbq}.

We wish in particular to reexamine one-sided constraints imposed on $\sigma_{{\rm DM}-X}/m_{\rm DM}$ by the Planck measurements of the CMB temperature and polarisation anisotropies in the light of prior dependence. 
 It is well known that parameter estimation and credible interval construction based upon Bayesian statistical inference are always subject to a greater or lesser extent to our choice of the prior probability distribution~\cite{Trotta:2017wnx}, i.e., the weights we assign to various parameter regions to quantify our beliefs about these regions {\it before} we look at the data.
 Even ``uninformative'' priors that have a uniform or ``flat'' probability density in the parameter directions of interest are not truly devoid of information; indeed, a uniform function in linear~$\theta$ represents a completely different probability density distribution to one that is flat in $\log \theta$, where the latter essentially corresponds to re-weighting linear $\theta$-space by the Jacobian determinant of the transformation, $|J| = 1/\theta$.

In this connection, we note that one-sided constraints are especially susceptible to poor choices of prior weight assignment in parameter regions to which the data have no sensitivity. Furthermore, the selection of prior boundaries in these regions where no natural, finite ones exist presents yet another vulnerability through which prior subjectivity may strongly influence the inference outcome.
This last point is especially pertinent to one-sided constraints on the dark matter--radiation coupling, where the natural lower boundary at $\sigma_{{\rm DM}-X}/m_{\rm DM}=0$ in linear $\sigma_{{\rm DM}-X}/m_{\rm DM}$-space becomes formally unbounded in $\log (\sigma_{{\rm DM}-X}/m_{\rm DM})$-space. Because the premise of Bayesian credible intervals construction is a finite total volume under the posterior probability distribution and credible intervals are but parameter regions containing the desired fractional volume, it is inevitable that any inferred upper limit on $\log (\sigma_{{\rm DM}-X}/m_{\rm DM})$ will depend to a large degree on the choice of proxy for $-\infty$.

It has come to our attention that a number of published upper limits on $\sigma_{{\rm DM}-X}/m_{\rm DM}$ in the recent literature
have been derived either under the assumption of a flat prior in $\log(\sigma_{{\rm DM}-X}/m_{\rm DM})$, or with no prior information specified. The issue of prior dependence and especially the pathology of the logarithmic flat prior in one-sided limits has been discussed in several cosmological contexts, e.g., primordial tensors~\cite{Valkenburg:2008cz} and neutrino masses~\cite{Heavens:2018adv}. Our purpose in this work is to raise awareness by reiterating the salient points in the context of CMB anisotropies constraints on dark matter--radiation interactions. We also take this opportunity to revise and/or update the Planck CMB constraints on $\sigma_{{\rm DM}-X}/m_{\rm DM}$ using better-behaved priors.

The paper is organised as follows. We begin in section \ref{Sec:Bayes} with a brief discussion of the various priors commonly employed in Bayesian statistical inference in the cosmological context, and detail in section \ref{sec:dmrad} the modelling and CMB phenomenology of dark matter--neutrino and dark matter--photon elastic scattering. In section \ref{Sec:Results} we test several dark matter--radiation scenarios against measurements of the CMB temperature and polarisation anisotropies by the Planck mission in a likelihood analysis 
 under various prior assumptions, and discuss our results therein.
 Section~\ref{Sec:Discussion} contains our conclusions.


\section{Priors in Bayesian statistical inference} \label{Sec:Bayes}

The basis of Bayesian statistical inference is Bayes's Theorem~\cite{Trotta:2017wnx},
\begin{equation}
\mathbb{P}\left(\theta|D\right) = \frac{\mathbb{P}(\theta) \, \,  \mathbb{P}\left(D|\theta\right)}{\mathbb{P}(D)} \equiv \frac{\mathbb{P}(\theta) \, \mathcal{L}(\theta) }{\int \, \mathbb{P}(\theta) \,  \mathcal{L}(\theta) \,\textrm{d}\theta}
\end{equation}
where $D$ denotes the data, and \(\theta \equiv \vec{\theta}\) represents the vector of model parameters subject to the analysis. 
After specifying the likelihood function~\(\mathcal{L}(\theta) \equiv \mathbb{P}(D|\theta)\), i.e., the probability of the data given the model parameters, and our prior beliefs on the probability distribution \(\mathbb{P}(\theta)\) of the parameters,
the goal is to find the posterior probability distribution~\(\mathbb{P}\left(\theta|D\right)\) of the parameters given the data. Once the posterior~\(\mathbb{P}\left(\theta|D\right)\) has been found, credible interval construction can proceed by way of identifying the parameter boundary surfaces that contain the desired fractional volume under~\(\mathbb{P}\left(\theta|D\right)\).

In the cosmological context, the accepted wisdom is that one should choose a prior distribution that is as uninformative as possible, so as to ``let the data decide'' where the parameter constraints should ultimately fall. Commonly used prior distributions include:

\paragraph{Linear flat prior} This simple choice corresponds to choosing a uniform distribution in the parameter direction $\theta$. However, because the volume of a uniform prior is strictly unbounded, in practical implementation (e.g., in a Markov Chain sampler) some prior boundaries need to be inserted by hand. However, as long as the likelihood function~${\cal L}(\theta)$ is sharply peaked within the prior region, the exact choice of these boundaries is generally immaterial as far as parameter estimation is concerned.  We do note however that there exist other statistical measures, such as the Bayesian evidence and the Kullback--Leibler divergence,  that are strongly dependent on the prior boundaries.  See, e.g.,~\cite{Trotta:2008qt}.

In the case where the data only have sufficient sensitivity to provide one-sided limits, fortunately, physically well-motivated parameters generally have physically well-motivated boundaries. In the context of dark matter--radiation interaction, the natural boundary $\sigma_{{\rm DM}-X}/m_{\rm DM}=0$ recovers the $\Lambda$CDM limit of no interaction. Furthermore, because a linear flat prior in $\theta$ naturally suppresses the weight assigned to a logarithmic interval of parameter space, ${\rm d} \ln \theta$, by a factor~$\theta$ as $\theta \to 0$, it avoids artificially enhancing the significance of parameter regions wherein the data cannot make any decision.

\paragraph{Logarithmic flat prior} When sampling a parameter direction that may vary over many orders of magnitude, it may be convenient to use a uniform prior in \(\log \theta \) to ensure that the parameter direction can be sampled efficiently. As with the linear flat prior, this can be a safe choice as long as the likelihood \(\mathcal{L}\) is sharply peaked within the prior region. An example in point is the primordial curvature power spectrum amplitude $A_s$, which is typically sampled as $\ln A_s$ in $\Lambda$CDM fits~\cite{Aghanim:2018eyx, Planck}.
 In some cases, adoption of a logarithmic flat prior may also be  justified on physical grounds,  e.g., the mass distribution of primordial black holes, in which case a logarithmic flat prior on the black hole mass simply reflects the generic expectations of the production mechanism~\cite{Clesse:2015wea,Chisholm:2005vm}.
 
However, where the likelihood function remains large at at least one prior boundary, then the choice of that boundary can significantly impact on credible interval construction. This is especially problematic in the context of constraining DM--radiation, for while $\sigma_{{\rm DM}-X}/m_{\rm DM}=0$ has a clear physical interpretation, its $\log(\sigma_{{\rm DM}-X}/m_{\rm DM})$ equivalent, i.e., $-\infty$, is ill-defined. A literal interpretation would call for picking a boundary that is as negative as is allowed by the sampling algorithm. However, at the same time, if the data do not have the requisite sensitivity to probe extremely small couplings, then employing a logarithmic flat prior on $\theta=\sigma_{{\rm DM}-X}/m_{\rm DM}$ amounts to artificially enhancing the volume of the unconstrainable parameter region by a factor $1/\theta$ relative to linear flat prior, while simultaneously suppressing the contribution from the large-$\theta$ region that {\it can} be probed by the data by the same factor. 

The impact on one-sided credible interval construction is immediately clear. Since by definition an $X$\%-credible interval is the parameter region that contains $X$\% of the volume under the posterior, any upper limit can be made arbitrarily tight by shifting the lower prior boundary to ever lower parameter values that cannot be distinguished by the data from a true $\sigma_{{\rm DM}-X}/m_{\rm DM}=0$ but nonetheless dominate the volume under the posterior solely through the pathology of the prior. This raises the question of whether one-sided limits derived using a logarithmic flat prior are actually meaningful.

\paragraph{Jeffreys prior} Defined as $\mathbb{P}(\theta) = \sqrt{|F(\theta)|}$, where \(|F(\theta)|\) denotes the determinant of the Fisher information matrix,
\begin{equation}
\label{eq:fisher}
F_{ij} = \mathbb{E}\left[\frac{\partial \ln \mathcal{L}}{\partial \theta_{i}} \frac{\partial \ln  \mathcal{L}}{\partial \theta_{j}}\right],
\end{equation}
and the operator \(\mathbb{E}[\cdots]\) the expectation value, the Jeffreys prior by construction maximises the effect of the data $D$ on the posterior \(\mathbb{P}\left(\theta|D\right)\)~\cite{10.2307/97883, 10.2307/2291752} and for this reason is often called an ``objective'' prior. For $n$ variables described by a multivariate Gaussian likelihood distribution,
\begin{equation}
\mathcal{L}(\theta)  = \frac{1}{\sqrt{(2 \pi)^n |M^{-1}|}} \exp\left[-\frac{1}{2}(\vec{q}-\vec{\theta}~)^{T}C^{{-{1} }}(\vec{q} - \vec{\theta}~)\right],
\end{equation}
where $C$ is the $n \times n$  ($\theta$-independent) covariance matrix, the Jeffreys prior evaluates to a uniform distribution \(\mathbb{P}(\theta) \propto 1/\sqrt{|C|}\).%
\footnote{As points out in~\cite{Cousins:2019cni}, this result applies independently of whether additional constraints should apply to the parameter values of $\vec{\theta}$, e.g., $\theta \geq 0$.}
In such cases,  credible intervals constructed under the Jeffreys prior are identical to those derived under a linear flat prior.
      For a general likelihood distribution, however, an explicit evaluation of the Fisher information matrix~(\ref{eq:fisher}) is required to establish the effects of the prior on parameter estimation.

Note that, though widely used in Bayesian analyses, the Jeffreys prior is strictly-speaking a non-Bayesian construct, in that its dependence on the Fisher information violates the likelihood principle by definition.
 In practice, this means that the prior probability distribution can vary even between two identical experiments testing the same model parameters under the same likelihood function.

\bigskip

We shall apply in the next sections the three types of prior distribution discussed above to analyse specific one-parameter extensions of the $\Lambda$CDM model involving dark matter--radiation elastic scattering.


\section{Dark matter--radiation scattering} \label{sec:dmrad}

We consider two scenarios in which the dark matter scatters elastically with (i) standard model neutrinos and (ii) photons. These scenarios have been previously explored in ~\cite{InteractingDM, InteractingDM2, LKD, LKD2, LKD3, Wilknu, Wilkphoton}.

\subsection{Dark matter--neutrino scattering} \label{Sec:DMnu}

Following~\cite{Wilknu}, the massless neutrino Boltzmann hierarchy in the conformal Newtonian gauge is modified to incorporate a DM--neutrino elastic scattering interaction as follows:
\begin{align}
\begin{split} \label{Eq:NeutrinoBoltzmann}
\dot{\delta}_{\nu} &= -\frac{4}{3}\theta_{\nu} + 4 \dot{\phi}, \\
\dot{\theta_{\nu}} &= k^{2}\psi + k^{2} \left(\frac{1}{4}\delta_{\nu} - \sigma_{\nu} \right) - \dot{\mu}_{\nu}\left(\theta_{\nu} - \theta_{\textrm{DM}}\right), \\
\dot{\sigma}_{\nu} &= \frac{4}{15}\theta_{\nu} - \frac{3}{10}k F_{\nu 3} - \frac{9}{10}\dot{\mu}_{\nu}\sigma_{\nu}, \\
\dot{F}_{\nu \ell} &= \frac{k}{2\ell+1}\left[\ell F_{\nu (\ell-1)} - \left(\ell + 1\right) F_{\nu (\ell + 1)}\right] - \dot{\mu}_{\nu} F_{\nu \ell}, \qquad \ell \geq 3,
\end{split}
\end{align}
where, using the convention of~\cite{MaB}, \(\delta_{\nu} = F_{\nu 0}, \, \theta_{\nu} = (3/4) k F_{\nu 1}\), and \(\sigma_{\nu} = F_{\nu 2}/2\) are the neutrino energy density, velocity divergence, and anisotropic stress respectively, \(F_{\nu \ell}\) is the \(\ell^{\textrm{th}}\) Legendre multipole moment, \(\theta_{\textrm{DM}}\) the DM velocity divergence, \(\psi\) and \(\phi\) the perturbations in the line element of the conformal Newtonian gauge \(\textrm{d}s^{2} = a^{2}[ -(1+2\psi)\textrm{d}\eta^{2} + (1-2\phi)\textrm{d}x^{i}\textrm{d}x_{j}]\), and an overdot denotes differentiation with respect to conformal time \(\eta\). The DM--neutrino conformal scattering rate is given by \(\dot{\mu}_{\nu} = a \sigma_{\textrm{DM}-\nu}n_{\textrm{DM}}\), where \(n_{\textrm{DM}}\) is the DM number density, and in writing equation~(\ref{Eq:NeutrinoBoltzmann}) we have implicitly assumed that the DM--neutrino interaction has the same angular dependance as Thomson scattering.

The corresponding equations of motion for dark matter perturbations are 
\begin{align}
\begin{split} \label{Eq:DMBoltzmann}
\dot{\delta}_{\textrm{DM}} &= - \theta_{\textrm{DM}} + 3 \dot{\phi}, \\
\dot{\theta}_{\textrm{DM}} &= k^{2} \psi - \mathcal{H} \theta_{\textrm{DM}} - S^{-1}_{\nu}\dot{\mu}_{\nu}\left(\theta_{\textrm{DM}}-\theta_{\nu}\right),
\end{split}
\end{align}
where the factor \(S_{\nu} = (3/4) \rho_{\textrm{DM}}/\rho_{\nu}\) is the ratio of the DM to neutrino energy densities, which arises from the conservation of momentum in the coupled DM--neutrino system. Note that we have omitted in the Euler equation a pressure gradient term proportional to the square of the intrinsic DM sound speed. Justification for this omission can be found in~\cite{Diacoumis:2017hff, Stadler:2018jin}.

\begin{figure}[t]
	\begin{center}
		\includegraphics[width=0.48\textwidth]{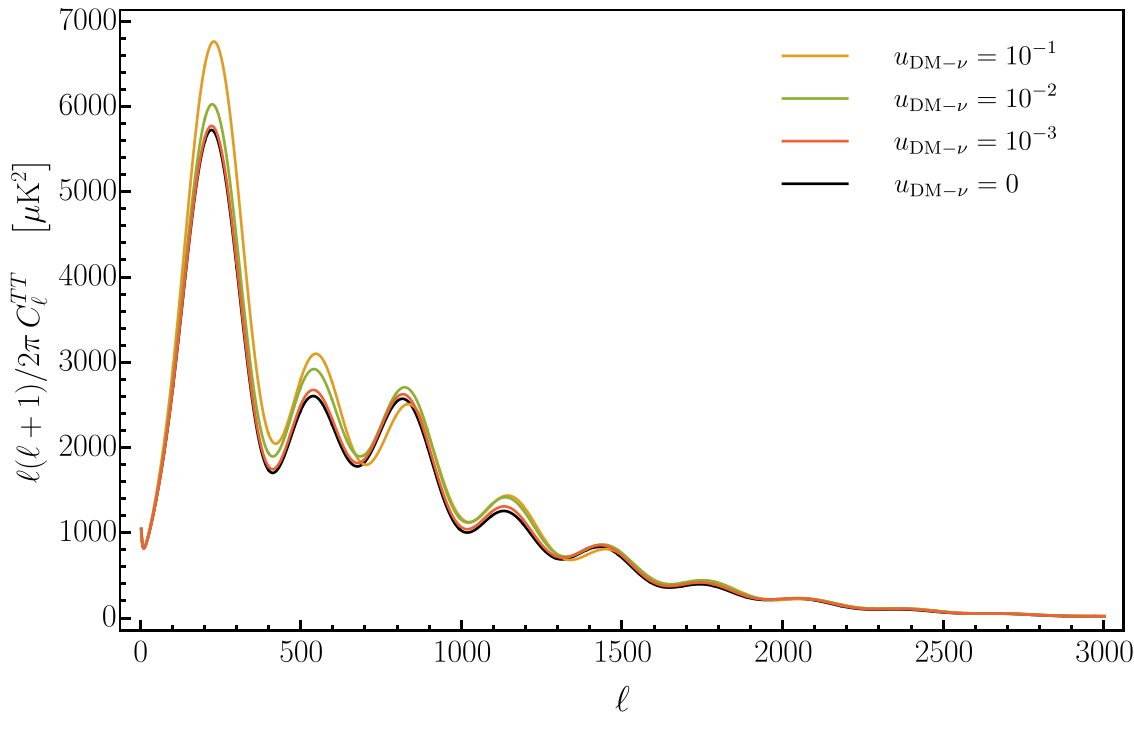}
		\includegraphics[width=0.48\textwidth]{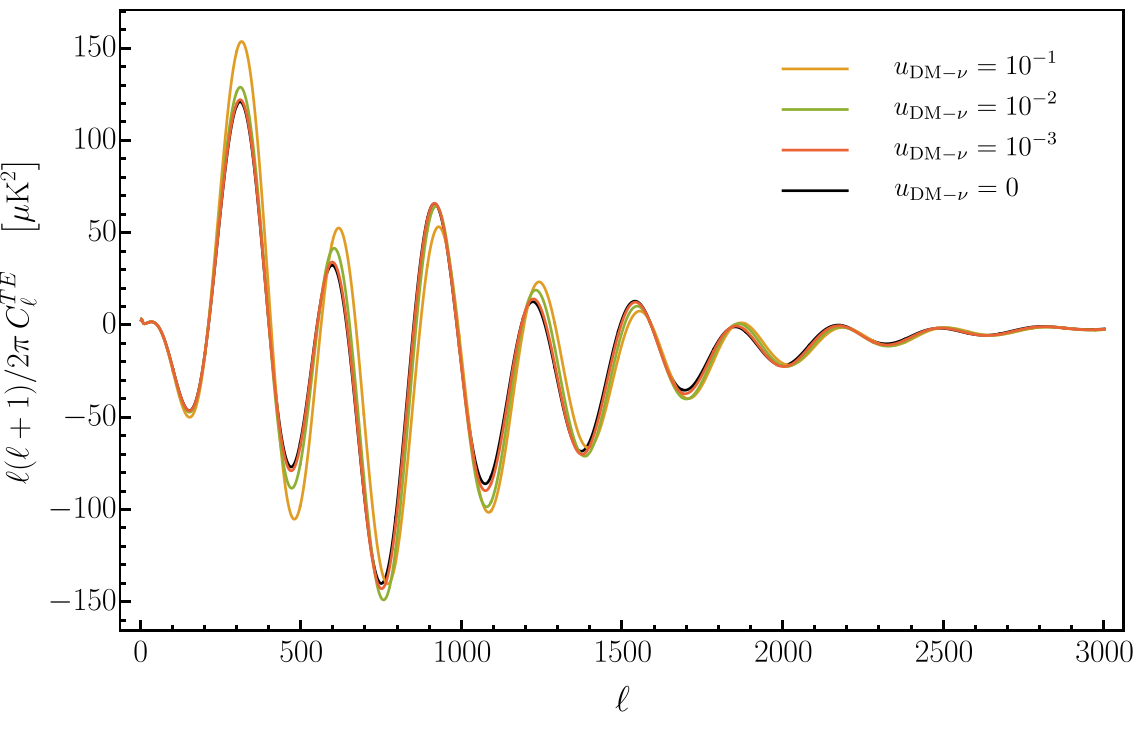}
		\includegraphics[width=0.48\textwidth]{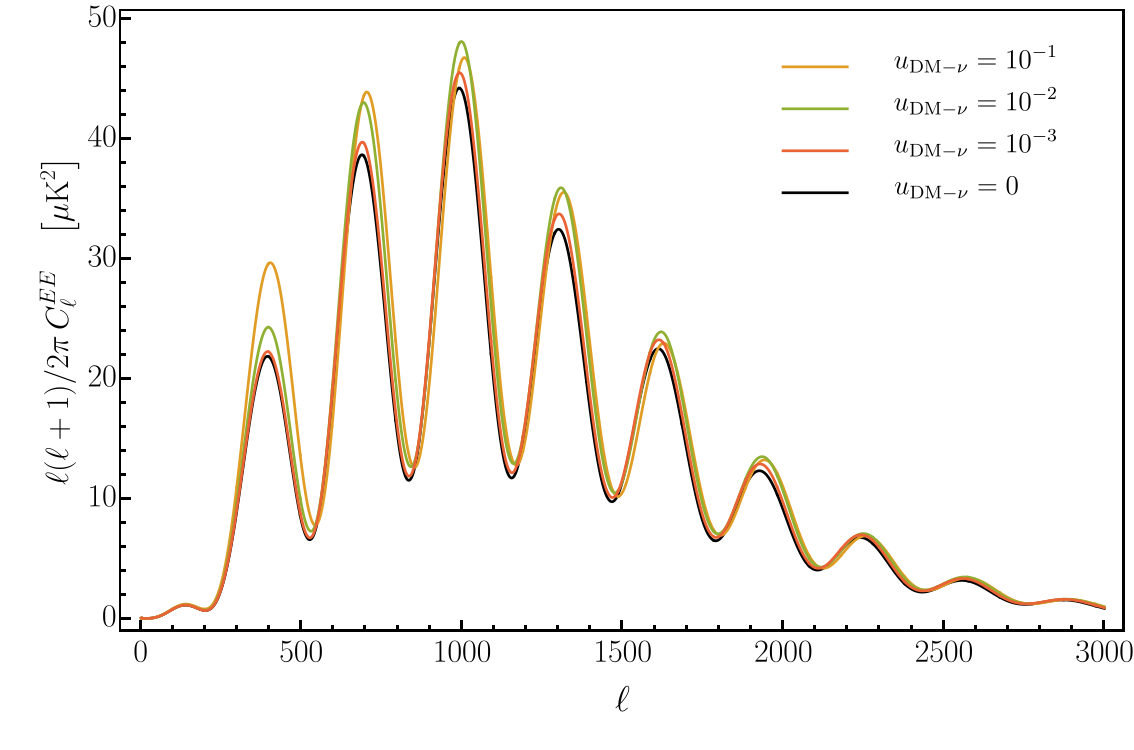}
		\caption{\textit{Top:} CMB temperature angular power spectrum in the presence of DM--neutrino interaction. The black, orange, green and red lines denote, respectively, \(u_{\nu} = 0, 10^{-1}, 10^{-2}\) and \(10^{-3}\), where \(u_{\nu}\) is defined in equation~(\ref{Eq:ufac}), and we have taken the elastic scattering cross-section to be constant in time. All other cosmological parameters have been set at their Planck 2015 best-fit values~\cite{Planck}. 
			\textit{Middle:} $TE$ cross correlation.
			\textit{Bottom:} $E$-mode polarisation.}
		\label{Fig:DMneutrinoCMB}
	\end{center}
\end{figure}

The specific effects of DM--neutrino interactions on the CMB have been discussed in detail in ~\cite{Wilknu}. The dominant effects can be understood in terms of the acoustic oscillations that develop when the dark matter and neutrinos form a tightly-coupled fluid. These oscillations are imprinted on the spacetime metric fluctuations, leading to observable consequences in the CMB temperature anisotropies, including (i) an increase in the acoustic peak heights at $\ell \gtrsim 200$ from the reduction of neutrino anisotropic stress, and (ii) a small shift in the positions of the acoustic peaks towards 
higher $\ell$ values, due to the DM--neutrino acoustic oscillations driving down the effective oscillation frequency of the gravitational potential at fixed wavenumbers.
Both of these effects can be clearly discerned in figure~\ref{Fig:DMneutrinoCMB}.

\subsection{Dark matter--photon scattering} \label{Sec:DMphot}

As in the case of DM--neutrino elastic scattering, DM--photon elastic scattering may be modelled via of a simple modification of the photon Boltzmann hierarchy by analogy with Thomson scattering. This amounts to introducing additional collision terms proportional to the conformal DM--photon scattering rate~$\dot{\mu}_\gamma=a \sigma_{{\rm DM}-\gamma} n_{\rm DM}$ on the r.h.s.\ of the photon Boltzmann equations. 

Specifically, the Boltzmann hierarchy for the photon temperature fluctuations becomes
\begin{align}
\begin{split}
\label{Eq:gammahierarchy}
\dot{\delta}_\gamma &= -\frac{4}{3} \theta_\gamma + 4 \dot{\phi}, \\
\dot{\theta}_{\gamma} &= k^{2}\psi + k^{2}\left(\frac{1}{4}\delta_{\gamma} - \sigma_{\gamma}\right)-\dot{\kappa} \left(\theta_{\gamma} - \theta_b\right) -\dot{\mu}_\gamma \left(\theta_{\gamma} - \theta_{\textrm{DM}}\right), \\
\dot{F}_{\gamma 2} & = 2\dot{\sigma}_{\gamma} =  \frac{8}{15}\theta_{\gamma}-\frac{3}{5}kF_{\gamma3}-\frac{9}{5}(\dot{\kappa} + \dot{\mu})\sigma_{\gamma}+\frac{1}{10}(\dot{\kappa} + \dot{\mu})\left(G_{\gamma 0} + G_{\gamma 2}\right), \\
\dot{F}_{\gamma \ell} & = \frac{k}{2 \ell +1} \left[ \ell F_{\gamma (\ell-1)} - (\ell +1) F_{\gamma (\ell+1)} \right] -  (\dot{\kappa}+ \dot{\mu}_\gamma) F_{\gamma \ell}, \qquad \ell \geq 3,
\end{split}
\end{align}
where we have used the notation of~\cite{MaB} in the conformal Newtonian gauge, and 
we identify $\delta_\gamma=F_{\gamma 0}$, \(\theta_{\gamma}=(3/4) k F_{\gamma 1}\), and $\sigma_\gamma= F_{\gamma 2}/2$ as the photon density perturbations, velocity divergence, and anisotropic stress respectively. The corresponding hierarchy for the photon polarisation fluctuations reads
\begin{align}
\label{Eq:polarisation}
\dot{G}_{\gamma \ell}  = \frac{k}{2 \ell +1} \left[ \ell G_{\gamma (\ell-1)} - (\ell +1) G_{\gamma (\ell+1)} \right]  + (\dot{\kappa}+ \dot{\mu}_\gamma) 
\left[- G_{\gamma \ell} +\frac{1}{2} \Pi \left( \delta_{\ell 0} + \frac{\delta_{\ell 2}}{5}\right)\right],
\end{align}
with $\Pi \equiv  F_{\gamma 2} + G_{\gamma 0} + G_{\gamma 2}$. Concurrently,
conservation of momentum in the coupled DM--photon system modifies the equation of motion for the dark matter perturbations to
\begin{align} \label{Eq:PhotonEuler}
\begin{split}
\dot{\delta}_{\rm DM} & = - \theta_{\rm DM} + 3 \dot{\phi}, \\
\dot{\theta}_{\textrm{DM}} &= k^{2}\psi - \mathcal{H}\theta_{\textrm{DM}} - S_{\gamma}^{-1}\dot{\mu}_\gamma \left(\theta_{\textrm{DM}} - \theta_{\gamma}\right) ,
\end{split}
\end{align}
where \(S_{\gamma} \equiv (3/4) \rho_{\textrm{DM}}/\rho_{\gamma}\), and we have again omitted in the Euler equation a pressure gradient term proportional to the square of the DM sound speed.

\begin{figure}[t]
	\begin{center}
		\includegraphics[width=0.48\textwidth]{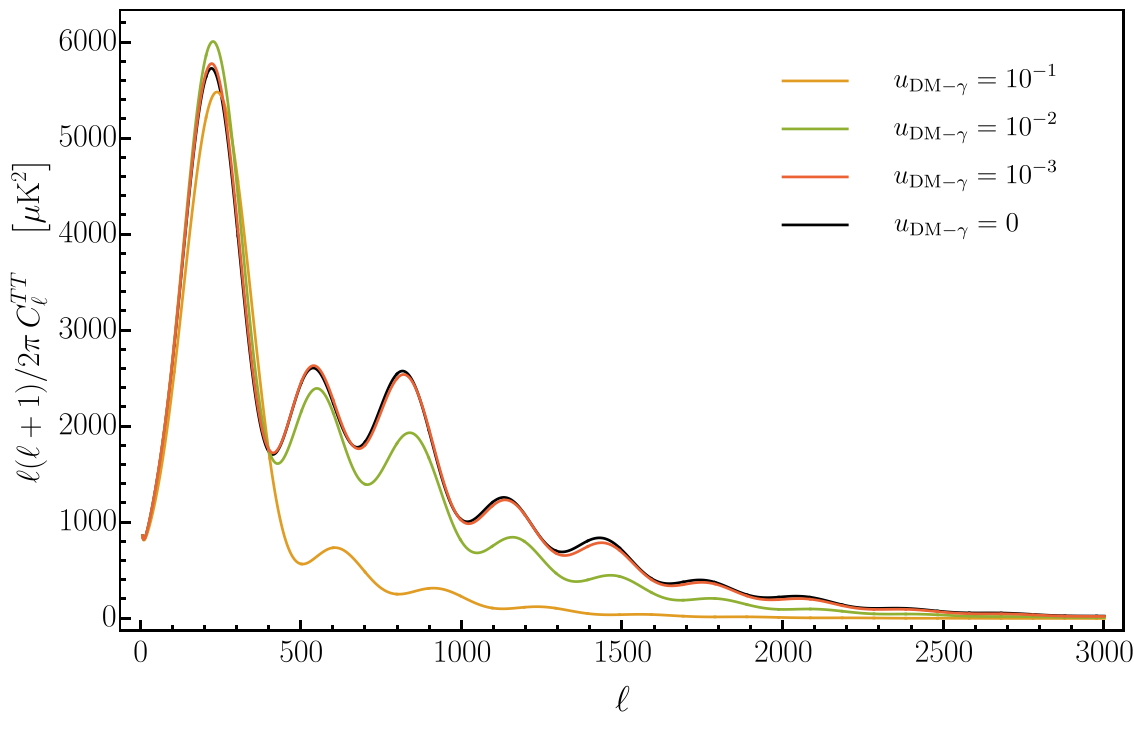}
		\includegraphics[width=0.48\textwidth]{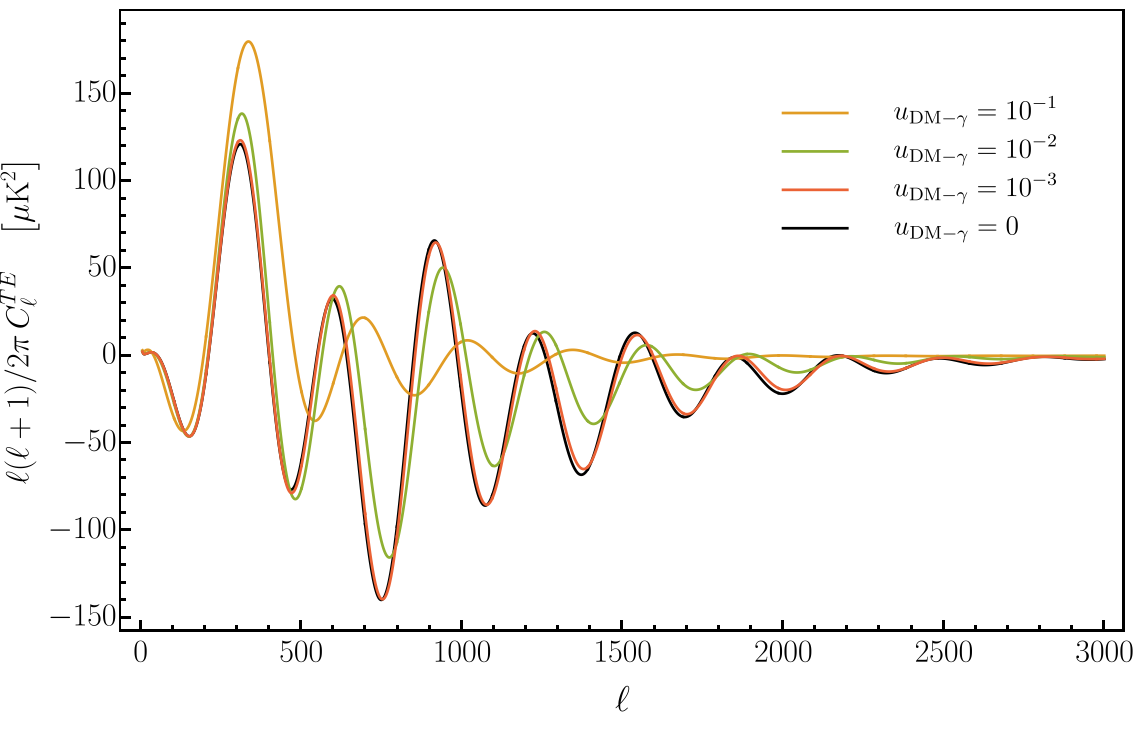}
		\includegraphics[width=0.48\textwidth]{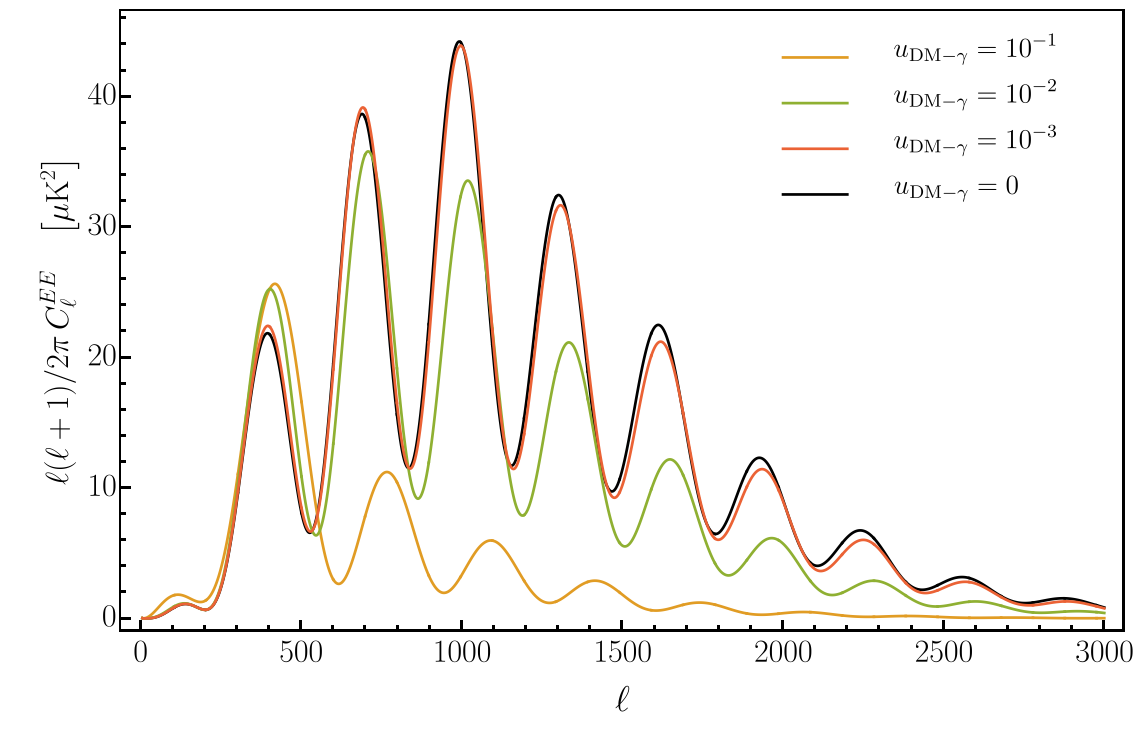}
		\caption{Same as figure~\ref{Fig:DMneutrinoCMB}, but for DM--photon interaction.}
		\label{Fig:DMphotonCMB}
	\end{center}
\end{figure}

As shown in figure \ref{Fig:DMphotonCMB}, DM--photon elastic scattering alters predominantly the damping tail of the CMB anisotropies through a modified diffusion damping scale $k_D$~\cite{Diacoumis:2017hff} :
\begin{equation} \label{Eq:SilkDampMod}
\partial_{z}k_{D}^{-2}(k) \simeq -\frac{c_{s}^{2}a}{2\mathcal{H}}\left[\frac{1}{\dot{\kappa} + \dot{\mu}_{\gamma}}\frac{16}{15} + \frac{3 \dot{\mu}_{\gamma}}{k^{2}}\left(\frac{k^{2}}{k^{2} + 3 S_{\gamma}^{-2}\dot{\mu}_{\gamma}^{2}}\right)\right],
\end{equation}
where
$c_s \simeq 1/\sqrt{3}$ is the photon--baryon fluid sound speed. Physically, the first term in equation~(\ref{Eq:SilkDampMod}) arises from viscosity damping, which is the dominant source of diffusion damping in $\Lambda$CDM; adding DM--photon scattering modifies the denominator as $\dot{\kappa} \to \dot{\kappa} + \dot{\mu}_\gamma$. The second term arises from heat conduction, which is always highly suppressed in $\Lambda$CDM, but can become the dominant mode of dissipation in a coupled DM--photon system at \(k \simeq S_{\gamma}^{-1} \dot{\mu}_{\gamma}\), when the DM transits from being strongly to weakly coupled to the photons and the ``slippage'' between the two fluids reaches a maximum. See~\cite{Diacoumis:2017hff} for a detailed discussion.

Besides diffusion damping, it is also evident in figure~\ref{Fig:DMphotonCMB} that the acoustic peak locations are shifted to higher $\ell$ values as a consequence of DM--photon scattering. This effect originates in a small correction to the coupled photon--baryon--DM fluid due to DM loading, which in turn lowers the frequency of the acoustic oscillations at fixed wavenumbers.


\section{Statistical inference} \label{Sec:Results}
\subsection{Model parameter space} 

We consider two one-parameter extensions to the standard six-variable \(\Lambda \textrm{CDM}\) fit, whose parameter spaces are spanned respectively by
\begin{equation}
\begin{aligned}
\label{eq:fitparams}
\vec{\theta}_\nu = \{\Omega_{b} h^{2}, \Omega_{c} h^{2}, H_0, \tau, n_s, \ln A_s, u_\nu^0 \}, \\
\vec{\theta}_\gamma = \{ \Omega_{b} h^{2}, \Omega_{c} h^{2}, H_0, \tau, n_s, \ln A_s, u_\gamma^0 \}, 
\end{aligned}
\end{equation}
where \(\Omega_{b} h^{2}\) is the physical baryon density parameter, \(\Omega_{c} h^{2}\) the physical cold dark matter density parameter, $H_0$ the present-day Hubble expansion rate, $\tau$~the optical depth to reionisation, and \(n_{s}\) and \(A_s\) denote respectively the spectral index and amplitude of the primordial curvature power spectrum. The quantities $u_\nu^0$ and $u_\gamma^0$ parameterises respectively the DM--neutrino and DM--photon elastic scattering, on which we elaborate below.

At the level of the Boltzmann hierarchies, we have seen in sections~\ref{Sec:DMnu} and~\ref{Sec:DMphot} that the sole quantities that describe DM--neutrino and DM--photon interactions are the conformal scattering rates $\dot{\mu}_X = a \sigma_{{\rm DM}-X} n_{\rm DM}$, where $X= \nu, \gamma$.  Rewriting these rates in terms of the standard fit parameter $\Omega_c h^2$, we find
\begin{equation}
\dot{\mu}_X  \simeq 1.0537 \times 10^{-5}  \ a^{-2} \ \Omega_c h^2 \ \frac{\sigma_{{\rm DM}-X}}{m_{\rm DM}}~{\rm GeV}\ {\rm cm}^{-3},
\end{equation}
where  the ratio $\sigma_{{\rm DM}-X}/m_{\rm DM}$ may now be identified as the new,  linearly independent variable in each extended fit.  Note that this identification alone does not imply that $\Omega_c h^2$ and  $\sigma_{{\rm DM}-X}/m_{\rm DM}$  are uncorrelated;   we can however expect them to be on physical grounds, since $\Omega_{c} h^{2}$ contributes to other well-measured effects---notably,  the odd CMB acoustic peak height ratios through their dependence on the epoch of matter--radiation equality---that cannot be mimicked by new interactions in the dark matter sector.

Following~\cite{Wilknu, Wilkphoton}, we represent the ratio $\sigma_{{\rm DM}-X}/m_{\rm DM}$  with the dimensionless parameter,
\begin{equation} \label{Eq:ufac}
u_{X} = \frac{\sigma_{\textrm{DM}-X}}{\sigma_{\rm T}} \, \left(\frac{100 \, \textrm{GeV}}{m_{\textrm{DM}}}\right),
\end{equation}
where \(\sigma_{\rm T} \simeq 6.65 \times 10^{-25}~{\rm cm}^2\) is the total Thomson scattering cross-section.
We consider cross-sections that scale with the temperature of the $X$ radiation as \(\sigma_{\textrm{DM}-X} \propto T_X^n\) for $n=0,2$. In terms of the $u_X$ parameter, this is equivalent to specifying \(u_X = u_X^{0} a^{-n}\), with \(u_X^{0}\) denoting its present-day value. 


\subsection{Data and analysis}
\label{sec:data}

We compute the CMB temperature and polarisation anisotropies for a range of parameter values in the parameter spaces~(\ref{eq:fitparams}), using the publicly available Boltzmann code \texttt{CLASS}~\cite{CLASSI, CLASSII, CLASSIV}\footnote{Available at \href{http://class-code.net/}{http://class-code.net/}} modified to include DM--neutrino scattering and DM--photon scattering as described in sections \ref{Sec:DMnu} and \ref{Sec:DMphot} respectively. We test these outputs against the Planck 2015 data using
\begin{enumerate}
\item the $TT$, $TE$, and $EE$ likelihood at \(\ell \geq 30\), 
\item the Planck low-\(\ell\) temperature+polarisation likelihood, and 
\item the Planck lensing likelihood,
\end{enumerate}
a combination referred to as ``\(TTTEEE\)+lowP+lensing'' in reference~\cite{Planck}. 
The parameter spaces of~(\ref{eq:fitparams}) are then sampled as Monte Carlo Markov Chains (MCMC) using the MCMC package~\texttt{MontePython}~\cite{Audren:2012wb}.\footnote{Available at \href{http://baudren.github.io/montepython.html}{http://baudren.github.io/montepython.html}} 

Note that for dark matter--neutrino interactions we take the neutrino mass sum to be identically zero \(\Sigma m_{\nu} = 0\), to ensure that the Boltzmann hierarchy~(\ref{Eq:NeutrinoBoltzmann}) applies to the entire neutrino population. For dark matter--photon interactions, on the other hand, we follow the Planck 2015 base \(\Lambda\textrm{CDM}\) analysis~\cite{Planck}, which assumes one massive and two massless neutrino species summing to \(\Sigma m_{\nu} = 0.06~\textrm{eV}\). In practice, however, the precise choices of \(\Sigma m_{\nu}\) and the mass spectrum have no statistically significant impact on the inference outcome as far as the Planck measurements are concerned.

\begin{table}[t]
	\begin{center}
		\begin{tabular}{|l|c|c|ccc|}
			\hline
			 & \multirow{2}{1.3cm}{Jeffreys} & \multirow{2}{1.8cm}{Linear flat} & &Logarithmic flat& \\ 
			 &&&1&2&3\\
			 \hline 
			$u_X^0, \; n=0$  &$[0, 0.1]$ &$[0, 0.1]$ & $[10^{-6}, 1]$ & $[10^{-5}, 1]$ & $[10^{-4}, 1]$\\ 
			$u_X^0, \; n=2$  &$ \left[0, 10^{-9}\right]$ & $\left[0, 10^{-9}\right]$ & $\left[10^{-15}, 10^{-11}\right]$ &$\left[10^{-14}, 10^{-11}\right]$&$\left[10^{-13}, 10^{-11}\right]$\\ 
			\hline 
		\end{tabular} 
		\caption{Prior boundaries for three types of priors imposed on the present-day values of the DM--$X$ coupling parameters $u_X^0$, where $X=\gamma, \nu$, and  \(u_{X} = u^{0}_{X}a^{-n}\).} \label{Tab:SumPrior}
	\end{center} 
\end{table}

The prior probability densities employed in the analysis for the variables of equation~(\ref{eq:fitparams}) are as follows:
\begin{itemize}
\item  For the $\Lambda$CDM variables $\Omega_{b} h^{2}, \Omega_{c} h^{2}, H_0, n_s$, and $\ln A_s$, 
we use improper linear flat priors unbounded at both ends.
\item For the optical depth to reionisation, we impose $\tau \in [0.01, \infty)$ following~\cite{Planck}.
\item  For the DM--$X$ coupling parameter $u_X^0$, we employ in turn the  Jeffreys prior, the linear flat prior, and three different logarithmic flat priors in the parameter regions specified in table~\ref{Tab:SumPrior}. 
\end{itemize}
 We estimate the Jeffreys prior in the $u^{0}_{X}$ direction from the 1D marginalised posterior distribution of $u^{0}_{X}$ constructed from the Markov Chains generated using a linear flat prior; in practice, this consists in taking the derivative of the logarithmic posterior with respect to $\theta=u^{0}_{X}$ as per equation~(\ref{eq:fisher}), where in this case the Fisher information matrix has only one (diagonal) entry.  The prior is then implemented for parameter estimation purposes by importance sampling the same Markov Chains.

As already discussed in section~\ref{Sec:Bayes}, while the linear flat and Jeffreys priors have a natural lower boundary at $u_X^0 =0$, there is no consistent and non-arbitrary way to cut off the logarithmic flat prior at the low end. Indeed, our choices of minima $u_X^0 = 10^{-6}$ ($n=0$) and $u_X^0 = 10^{-15}$($n=2$) are motivated by no deeper physical reason than mere convenience---lowering these boundaries further significantly increases the amount of time it takes for the Markov Chains to converge, as the MCMC sampler wastes time exploring large swaths of high-likelihood parameter space to which the data have no sensitivity. Updating the prior {\it after} having seen the data likelihood function is of course precisely what one should {\it not} be doing in a Bayesian statistical inference. That we have had to resort to this trick represents yet another pathology of the logarithmic flat prior for one-sided constraints.

Lastly, we note that the upper prior boundaries in table~\ref{Tab:SumPrior} have been chosen in the interest of preserving the stability of \texttt{CLASS}. Very large values of \(u_X^0\) cause the equations of motion~(\ref{Eq:NeutrinoBoltzmann}) and (\ref{Eq:PhotonEuler}) to become stiff; the usual workaround is to employ the tightly-coupled approximation for the dark matter and radiation fluid~\cite{Chacko:2016kgg}. For simplicity, we opt not to implement this approximation and cut off the parameter region where stiffness is expected. As we shall see, this choice has no serious impact on our parameter constraints, as the posterior distributions in $u_X^0$ always tend to zero well before reaching the upper prior boundary.


\subsection{Results and discussions}

\begin{figure}[t]
	\begin{center}
\includegraphics[width=15cm]{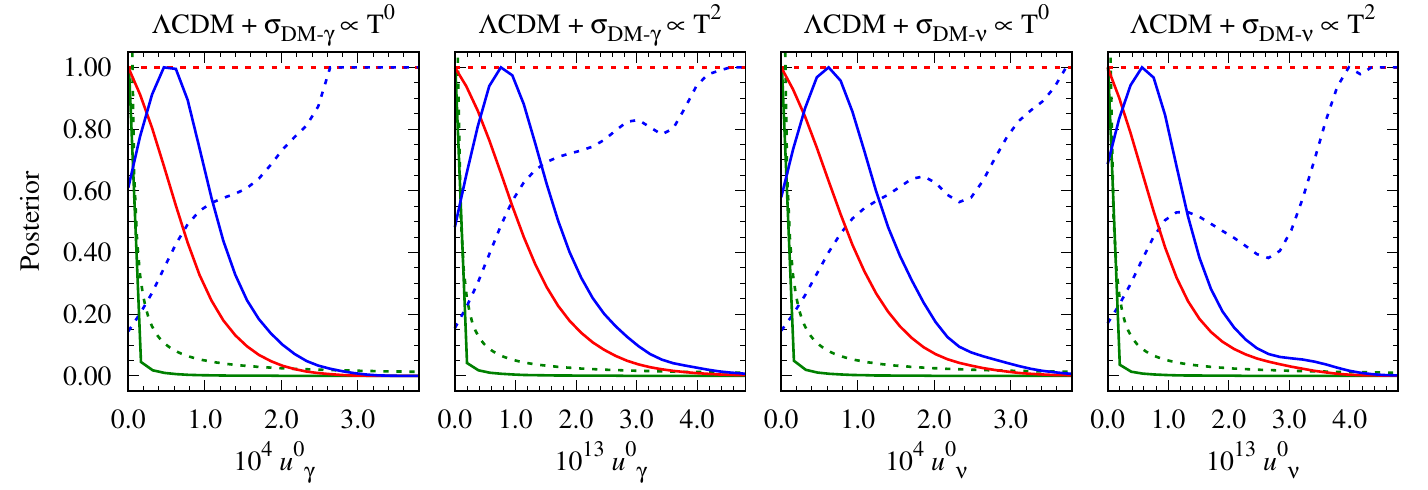}
\includegraphics[width=15cm]{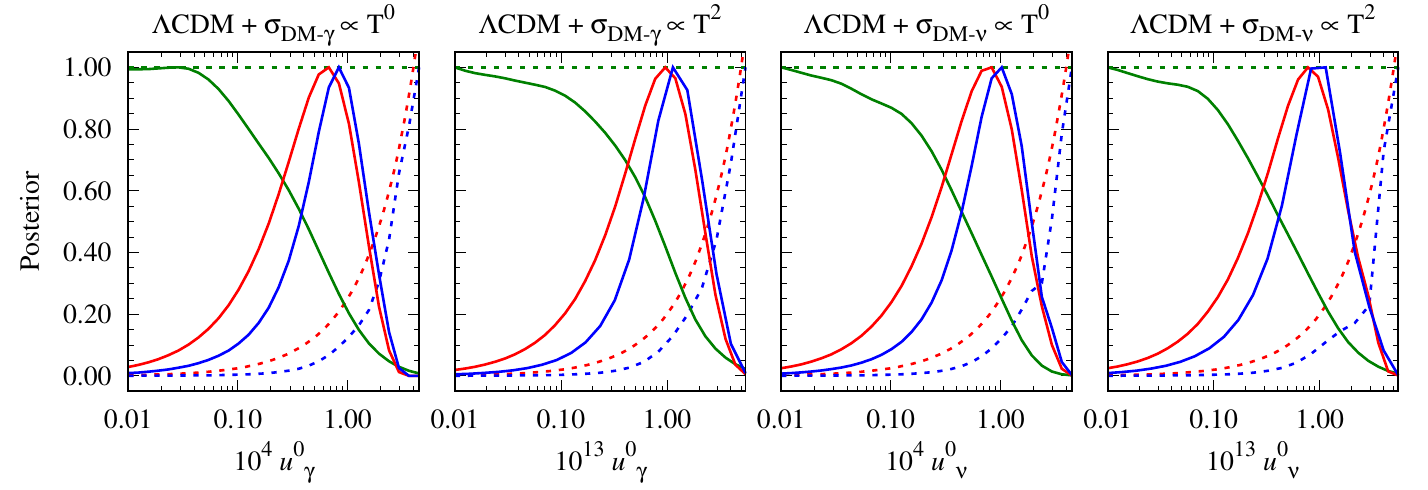}
		\caption{1D marginalised posterior distributions of the DM--$X$ elastic coupling parameter \(u^{0}_X\) (solid lines), derived from the Planck 2015 data assuming the  Jeffreys prior (blue), linear flat prior (red), and logarithmic flat prior 1 (green) on \(u^{0}_X\).   The prior distributions themselves are denoted by the dotted lines.
{\it Top row}: The distributions represented on a linear scale.  {\it Bottom row}: The same distributions represented on a logarithmic scale.}
\label{Fig:MCMCgconst}
	\end{center}
\end{figure}

Figure~\ref{Fig:MCMCgconst} shows the 1D marginalised posterior distributions of the DM--radiation coupling parameters $u_X^0$, assuming in turn the Jeffreys prior, linear flat prior, and logarithmic flat prior~1 on $u_X^0$, in the four scenarios considered in this work.   For completeness the prior distributions themselves are also shown in the figure.%
\footnote{Estimates of the Jeffreys prior from Markov Chains are generally poor and noisy at the high end of  $u_X^0$ where samples are scarce.  However, because the posterior distribution is correspondingly small at the tail,  the exact behaviour of the prior in this limit has a negligible impact on parameter estimation.  For aesthetics we have shown in figure~\ref{Fig:MCMCgconst} our estimates of the Jeffreys prior at  $u_X^0 \leq u_{X,{\rm min}}^0$ and substituted it with a uniform distribtuion at $u_X^0 > u_{X,{\rm min}}^0$, where $u_{X,{\rm min}}^0$ is the lowest value of $u_X^0$ for which this procedure yields no more than a 0.1\% change in the 95\% limits on $u_X^0$.}
 The corresponding 95\% C.L.~upper limits on $u_X^0$ are summarised in table~\ref{Tab:uvals}.

In all four cases, we find that relative to the linear flat prior, the role of the  Jeffreys prior is to ``de-weight'' the posterior distribution in the peak region close to $u_X^0=0$, while the logarithmic flat prior plays the opposite role of de-weighting the tail region. Consequently, as shown in table~\ref{Tab:uvals}, the Jeffreys prior typically yields the weakest 95\%~C.L.~upper limits on $u_X^0$, followed by the linear flat prior for which the limits are some $10\to 20$\% tighter.
Imposing logarithmic flat prior~1 tightens the bounds by another factor of about $2 \to2.5$.
Note that the choice of prior on $u_X^0$ has no statistically significant impact on the other six parameter constraints,  nor is $u_X^0$ correlated with any other parameter.
The interested reader can find the full set of 1D marginalised constraints in appendix~\ref{Sec:Tables}.

\begin{table}[t]
	\begin{center}
		\begin{tabular}{|l|c|c|ccc|}
			\hline
			\multirow{2}{2cm}{Parameter} & \multirow{2}{1.3cm}{Jeffreys} & \multirow{2}{1.8cm}{Linear flat} & &Logarithmic flat &\\
			&&&1&2& 3\\ \hline 
			$10^{4}\, u^{0}_{\gamma}, \;\;\, n=0$  &  $<1.719 $ &  $<1.442$ & $<0.709$&$<1.171$&$< 2.480$\\ 
			$10^{13}\, u^{0}_{\gamma},  \; n=2$  & $<2.735 $ & $<2.320$ & $<1.070$ & $<1.552$ &$<3.500$\\ 
			$10^{4}\, u^{0}_{\nu},  \;\;\, n=0$ & $<2.141$ & $<1.810~$ & $<0.735$ & $< 1.034$&$ <1.740$\\ 
			$10^{13}\, u^{0}_{\nu}, \; n=2$ & $< 2.463$ &  $< 2.190$ & $<0.839$& $<1.293$ & $<2.775$\\ 
			\hline 
		\end{tabular} 
		\caption{1D marginalised \(95\%\) confidence limits on \(u^{0}_{X}\) corresponding to figure~\ref{Fig:MCMCgconst}.} \label{Tab:uvals}
	\end{center} 
\end{table}

 Observe also in table~\ref{Tab:uvals} that the bounds derived under a logarithmic flat prior are strongly dependent on our choice of lower boundaries.  In all four cases, moving the lower boundary up by two orders of magnitude (i.e., switching from logarithmic flat prior 1 to prior 3) relaxes the bound on $u_X^0$ by a factor of $2.5 \to 3$.  This can be understood from an inspection of the bottom row of figure~\ref{Fig:MCMCgconst}, where it is immediately clear that shifting the lower prior boundary up corresponds to excluding from credible interval construction a large volume of parameter space at which the 1D posterior distribution derived under a logarithmic flat prior is essentially flat and at its maximum.   Similarly, if we were to shift the prior boundary down, we could expect the same kind of volume effects to act  to tighten the bound on $u_X^0$.  Since there is no guideline to choosing the correct lower prior boundary, this exercise therefore highlights the arbitrariness of one-sided limits derived under a logarithmic flat prior.

A survey of existing CMB anisotropies bounds in the recent literature on dark matter--radiation elastic scattering is presented in table~\ref{Tab:Litvals}. Some remarks are in order:
\begin{enumerate}
\item The same group of authors covered in~\cite{Wilknu, Wilkphoton} all four cases. However, absent prior assumption and fitting the older P13 dataset (Planck 2013 high-\(\ell\) and low-\(\ell\) temperature+WMAP low-\(\ell\) polarisation),
their results cannot be easily compared with ours, especially as the former have been presented as one-sided \(68\%\) confidence limits, instead of the more conventional choice of 95\%.

Nonetheless we note that, order-of-magnitude-wise, there is reasonable agreement to within factors of 2.5, 
except in the case of \(n = 0\) DM--neutrino scattering, where the limit $10^4\, u_\nu^0 < 399~(68\%)$ from~\cite{Wilknu} 
 differs from our $10^4\, u_\nu^0< 1.810~(95\%)$ by a full two orders of magnitude. In view that improvements in parameter constraints have generally been moderate across the Planck 2013 and 2015 data releases~\cite{Planck}, we are inclined to conclude that the $u_\nu^0$ $(n=0)$ bound of~\cite{Wilknu} is erroneous. The result of~\cite{Escudero:2015yka} corroborates this conclusion. See point 2 below.

\item Reference~\cite{Escudero:2015yka} explored DM--neutrino scattering, and as in~\cite{Wilknu, Wilkphoton}, employed the P13 dataset. In the $n=0$ case, their choice of a logarithmic flat prior on \(u^{0}_{\nu} \in \left[10^{-6}, 1 \right]\) is 
 identical to ours (see table~\ref{Tab:SumPrior}). Expectedly, the corresponding bound from our analysis, $10^4\, u_\nu^0 < 0.735~(95\%)$, is 
only marginally tighter than their $10^4\, u_\nu^0< 0.912~(95\%)$, where the 20\% difference is likely attributable to improved polarisation measurements by Planck over WMAP. 

 In reference to point 1 above, this technical agreement supports our conclusion that $u_\nu^0$ $(n=0)$ bound of~\cite{Wilknu} is incorrect. Notwithstanding, we emphasise that the logarithmic bounds of both~\cite{Escudero:2015yka} and our analysis are a factor of 2.5 more stringent than those derived under the  Jeffreys and the linear flat prior, $10^4\, u_\nu^0\lesssim 1.8 \to 2.1~(95\%)$, for no better reason than as an artefact of excessive weight assignment in the low $u_\nu^0$ region. These bounds, therefore, cannot be but treated with suspicion.
 
In the \(n=2\) case, the choice of a logarithmic flat prior on \(u^{0}_{\nu} \in \left[10^{-18}, 10^{-11} \right]\) in~\cite{Escudero:2015yka} amounts to shifting the lower prior boundary down by three orders of magnitude relative to our settings (table~\ref{Tab:SumPrior}). Not surprisingly, this leads to an upper bound $10^{13}\, u^{0}_{\nu} < 0.251~(95\%)$ that is tighter than our $10^{13}\, u^{0}_{\nu} < 0.839~(95\%)$
by a factor of \(3.3\). Relative to our $10^{13}\, u^{0}_{\nu} < 2.463~(95\%)$  derived under the Jeffreys prior, the difference is an order of magnitude. This again highlights the pathology of the logarithmic flat prior, where any one-sided limit can be made arbitrarily tight by shifting the lower prior boundary. 

\item Reference~\cite{Stadler:2018jin} analysed the case of DM--photon scattering assuming \(n = 0\), deriving constraint on $u_\gamma^0$ using identically the same P15 dataset used in our analysis. Indeed, what distinguishes their analysis from ours is the inclusion of several small corrections in the equations of motion, such as the dark matter sound speed and the tight-coupling approximation at second order. 
No prior assumption on $u^{0}_{\gamma}$ has been specified in~\cite{Stadler:2018jin}, but the bound $10^{4}\, u^{0}_{\gamma} < 1.490~(95\%)$ is in excellent agreement with our $10^{4}\, u^{0}_{\gamma} < 1.442~(95\%)$ under a linear flat prior.

\item Reference~\cite{DiValentino:2017oaw} tested the case of \(n=0\) DM--neutrino scattering, again against the P15 dataset.
Using a logarithmic flat prior in \(u^{0}_{\nu}\)---but without specifying explicitly the prior boundaries---they obtained $10^4\, u_\nu^0< 0.794~(95\%)$ that is in remarkably good agreement with our $10^4\, u_\nu^0< 0.735~(95\%)$ under a logarithmic flat prior. We can only surmise here that a very similar prior range must have been used in~\cite{DiValentino:2017oaw}. Whatever the answer, these bounds are a factor 2.5 more stringent than they need to be, and should be disregarded in favour of those derived under either the Jeffreys or the linear flat prior in table~\ref{Tab:uvals}.

\end{enumerate}

\begin{table}[t]
	\begin{center}
		\begin{tabular}{|l|c|c|c|c|c|}
			\hline
			Parameter & Upper limit & Prior type & Prior boundaries & Data \\ \hline 
			$10^{4}\, u^{0}_{\gamma}, \;\;\, n=0$  & $<1.173~(68\%)$~\cite{Wilkphoton} & Unspecified  & Unspecified & P13+WP \\ 
			& $<1.490~(95\%)$~\cite{Stadler:2018jin} & Unspecified & Unspecified & P15\\ 
			\hline
			$10^{13}\, u^{0}_{\gamma}, \; n=2$  & $<0.9\, (68\%)$~\cite{Wilkphoton}  & Unspecified & Unspecified & P13+WP \\ \hline
			$10^{4}\, u^{0}_{\nu},  \;\;\, n=0$ & $< 399~(68\%)$~\cite{Wilknu} &  Unspecified & Unspecified & P13+WP \\
			& $< 0.912~(95\%)$~\cite{Escudero:2015yka} & Log flat & $[10^{-6},1]$ & P13+WP\\
			& $< 0.794~(95\%)$~\cite{DiValentino:2017oaw}& Log flat & Unspecified & P15 \\  \hline
			$10^{13}\, u^{0}_{\nu}, \; n=2$ & $<2.56 \, (68\%)$~\cite{Wilknu} &Unspecified & Unspecified & P13+WP\\
			& $< 0.251~(95\%)$~\cite{Escudero:2015yka}& Log flat & $[10^{-18},10^{-11}]$ & P13+WP\\ 
			\hline 
		\end{tabular} 
		\caption{1D marginalised \(95\%\) confidence limits on \(u^{0}_{X}\), together with their corresponding prior assumption and data input, in the recent literature. The shorthand ``P13+WP'' denotes Planck 2013 high-\(\ell\) and low-\(\ell\) temperature combined with WMAP low-\(\ell\) polarisation measurements; ``P15'' is identically the Planck 2015 $TTTEEE$+lowP+lensing dataset used in our analysis, a description of which can be found in section~\ref{sec:data}.} \label{Tab:Litvals}
	\end{center} 
\end{table}


\noindent In summary, we advocate as our ``objective'' 95\% confidence limits the numbers in table~\ref{Tab:uvals} computed from the Planck 2015 \(TTTEEE\)+lowP+lensing data combination under the assumption of the  Jeffreys prior. Translated using equation~(\ref{Eq:ufac}) into constraints on the elastic scattering cross-sections, these are equivalently
 \begin{equation}
\begin{aligned}
\label{eq:constraintsgamma}
&\sigma_{\rm DM-\gamma} < 1.72 \times 10^{-6} \, \sigma_{\rm T} \left(\frac{m_{\rm DM}}{{\rm GeV}} \right) \simeq 1.14 \times 10^{-30} \,  \left(\frac{m_{\rm DM}}{{\rm GeV}} \right)~{\rm cm}^2
,  \quad n=0 \\
&\sigma_{\rm DM-\gamma}  <  2.74 \times 10^{-15} \, \sigma_{\rm T} \left(\frac{m_{\rm DM}}{{\rm GeV}} \right)
 \simeq 1.82 \times 10^{-39} \,  \left(\frac{m_{\rm DM}}{{\rm GeV}} \right)~{\rm cm}^2,  \quad n=2
\end{aligned}
\end{equation}
for dark matter--photon interaction, and
\begin{equation}
\begin{aligned}
\label{eq:constraintsnu}
&\sigma_{\rm DM-\nu} < 2.14 \times 10^{-6} \, \sigma_{\rm T} \left(\frac{m_{\rm DM}}{{\rm GeV}} \right) \simeq 1.42 \times 10^{-30} \,  \left(\frac{m_{\rm DM}}{{\rm GeV}} \right)~{\rm cm}^2, \quad  n=0 \\
&\sigma_{\rm DM-\nu}< 2.46 \times 10^{-15} \, \sigma_{\rm T} \left(\frac{m_{\rm DM}}{{\rm GeV}} \right) \simeq 1.64 \times 10^{-39} \,  \left(\frac{m_{\rm DM}}{{\rm GeV}} \right)~{\rm cm}^2, \quad  n=2
\end{aligned}
\end{equation}
for their dark matter--neutrino counterpart.


\section{Conclusions} \label{Sec:Discussion}

In this work, we have examined the prior dependence of one-sided limits on dark matter--radiation elastic scattering derived from CMB temperature and anisotropies measurements, and in so doing presented in table~\ref{Tab:uvals} a new set of constraints on the coupling parameters/interaction cross-sections computed from the 2015 data of the Planck mission.
We have tested in particular the linear flat, logarithmic flat, and  Jeffreys priors in cases where the dark matter scatters elastically with neutrinos or photons, assuming two different time dependences for the scattering cross-sections.
 
We find that in all four cases constraints derived under the linear flat and the Jeffreys prior agree with one another to within 20\%, with the latter bounds being in general the laxer ones.  In contrast, those computed with a logarithmic flat prior in our exercise are typically a factor of $2 \to 3$ too tight. 
Indeed, we have explicitly demonstrated that  the logarithmic flat prior can be highly pathological in that one-sided limits can always be made artificially tight simply by adjusting the lower prior boundary to ever lower parameter values to which the data have no sensitivity. Given this pathology, we question whether one-sided constraints derived under the logarithmic flat prior, though technically well defined, can ever have meaningful objective interpretations.

Regrettably, surveying the recent literature, we find that several existing constraints in the context of dark matter--neutrino interaction have been derived using the logarithmic flat prior on the scattering cross-section, resulting in upper bounds that are up to an order of magnitude (artificially) tighter than they ought to objectively be. Even more works have  omitted to specify their prior assumptions altogether, which is certainly not good practice in any statistical analysis.

Our revised ``objective'' constraints on the dark matter--photon and dark matter--neutrino elastic scattering cross-sections from the Planck 2015 temperature and polarisation measurements, given in equations~(\ref{eq:constraintsgamma}) and~(\ref{eq:constraintsnu}), 
have been computed under the assumption of the  Jeffreys prior. To minimise confusion in the future, we strongly urge all authors to accord more attention to prior dependence in the computation of one-sided limits.


\acknowledgments

JADD acknowledges support from an Australian Government Research
Training Program Scholarship. The work of Y$^3$W is supported in part by the Australian
Government through the Australian Research Council's Discovery Project funding scheme
(project DP170102382).



\appendix
\section{Parameter estimates tables} \label{Sec:Tables}

Tables~\ref{Tab:MCMCgconst} to~\ref{Tab:MCMCnuT2} show the mean values and 1D marginalised \(68\%\) credible intervals for the base $\Lambda$CDM variables of equation~(\ref{eq:fitparams}), together with  the 1D marginalised 95\% confidence limits on the coupling parameters $u_X = u_X^0 a^{-n}$, where $X=\gamma, \nu$ and $n=0,2$.
All have been derived from the Planck 2015 $TTTEEE$+lowP+lensing data combination with a pivot scale of \(k_{*} = 0.05~\textrm{Mpc}^{-1}\).

\bigskip
\begin{table}[h]
	\begin{center}
	\begin{tabular}{|l|c|c|c|}
		\hline
		Parameter  &  Jeffreys & Linear flat & Logarithmic flat 1\\ \hline 
		$10^{4}\, u^{0}_{\gamma}, \; n=0$  & $<1.719~(95 \%) $ & $<1.442~(95 \%)$ & $<0.709~(95 \%)$\\ 
		$100~\omega_{b }$ & $2.228_{-0.016}^{+0.016}$  & $2.227_{-0.016}^{+0.015}$ & $2.226_{-0.017}^{+0.015}$\\ 
		$\omega_{\textrm{cdm}}$  & $0.1199_{-0.0015}^{+0.0014}$  & $0.1197_{-0.0015}^{+0.0014}$ & $0.1195_{-0.0015}^{+0.0014}$ \\ 
		$100~\theta_{s }$  & $1.042_{-0.00036}^{+0.00040}$   & $1.042_{-0.00039}^{+0.00035}$ & $1.042_{-0.00034}^{+0.00031}$ \\
		 $\ln\left(10^{10}A_{s }\right)$ & $3.067_{-0.026}^{+0.025}$  & $3.066_{-0.026}^{+0.025}$ & $3.064_{-0.025}^{+0.026}$ \\ 
		$n_{s }$  & $0.9644_{-0.0049}^{+0.0047}$ & $0.9644_{-0.005}^{+0.0047}$ & $0.9646_{-0.0048}^{+0.0049}$\\ 
		$\tau_{\textrm{reio}}$ & $0.06661_{-0.014}^{+0.014}$ & $0.06625_{-0.014}^{+0.014}$ & $0.06573_{-0.014}^{+0.014}$\\ 
		$H_{0}$  & $67.47_{-0.64}^{+0.66}$  & $67.47_{-0.64}^{+0.66}$ & $67.49_{-0.65}^{+0.66}$ \\
		\hline 
	\end{tabular} 
	\caption{Dark matter--photon coupling; time-independent scattering cross-section.}
	\label{Tab:MCMCgconst}
\end{center} 
\end{table}

\bigskip

\begin{table}[h!]
	\begin{center}
	\begin{tabular}{|l|c|c|c|} 
		\hline
		Parameter  & Jeffreys & Linear flat & Logarithmic flat 1\\ \hline 
		$10^{13}\, u^{0}_{\gamma}, \; n=2$  & $<2.735~(95 \%) $ & $<2.320~(95 \%)$ & $<1.070~(95 \%)$ \\ 
		$100~\omega_{b}$ & $2.219_{-0.016}^{+0.016}$ & $2.220_{-0.016}^{+0.016}$ & $2.223_{-0.016}^{+0.016}$\\
		$\omega_{\textrm{cdm}}$ & $0.1197_{-0.0014}^{+0.0015}$  & $0.1197_{-0.0014}^{+0.0015}$ & $0.1195_{-0.0015}^{+0.0015}$ \\ 
		$100~\theta_{s}$  & $1.042_{-0.00031}^{+0.00030}$ & $1.042_{-0.00031}^{+0.00031}$ & $1.042_{-0.00030}^{+0.00030}$\\ 
		$\ln\left(10^{10}A_{s}\right)$  & $3.058_{-0.025}^{+0.024}$  & $3.059_{-0.025}^{+0.025}$ & $3.062_{-0.025}^{+0.026}$ \\ 
		$n_{s}$  & $0.9588_{-0.0059}^{+0.0054}$  & $0.9601_{-0.0053}^{+0.0058}$ & $0.9629_{-0.0051}^{+0.0055}$ \\ 
		$\tau_{\textrm{reio}}$ & $0.0634_{-0.014}^{+0.013}$  & $0.0639_{-0.013}^{+0.013}$ & $0.06487_{-0.014}^{+0.014}$ \\ 
		$H_{0}$  & $67.35_{-0.68}^{+0.62}$  & $67.35_{-0.66}^{+0.64}$ & $67.43_{-0.66}^{+0.65}$ \\ 
		\hline 
	\end{tabular}
	\caption{Dark matter--photon coupling; scattering cross-section scales as $\propto T^2$.} \label{Tab:MCMCgT2}
    \end{center}
\end{table}

\bigskip

\begin{table}[h!]
	\begin{center}
\begin{tabular}{|l|c|c|c|c|} 
	\hline 
	Parameter  &  Jeffreys & Linear flat & Logarithmic flat 1\\ \hline 
	$10^{4}\, u^{0}_{\nu}, \; n=0 $ & $<2.141~(95 \%)$ & $<1.810~(95 \%)$ & $<0.735~(95 \%)$\\ 
	$100~\omega_{b }$ & $2.226_{-0.016}^{+0.016}$ & $2.226_{-0.016}^{+0.016}$ & $2.227_{-0.016}^{+0.017}$ \\ 
	$\omega_{\textrm{cdm}}$ &  $0.1197_{-0.0014}^{+0.0015}$ & $0.1196_{-0.0015}^{+0.0015}$ & $0.1193_{-0.0015}^{+0.0015}$ \\ 
	$100~\theta_{s }$ & $1.042_{-0.00030}^{+0.00031}$ & $1.042_{-0.00030}^{+0.00030}$ & $1.042_{-0.00033}^{+0.00032}$ \\ 
	$\ln\left(10^{10}A_{s}\right)$ & $3.053_{-0.026}^{+0.025}$ & $3.053_{-0.025}^{+0.025}$ & $3.055_{-0.026}^{+0.027}$ \\ 
	$n_{s}$ & $0.9615_{-0.0057}^{+0.0050}$ & $0.9624_{-0.0053}^{+0.0053}$ & $0.9644_{-0.0054}^{+0.0051}$ \\ 
	$\tau_{\textrm{reio}}$ & $0.0606_{-0.014}^{+0.014}$ & $0.06082_{-0.014}^{+0.014}$ & $0.06157_{-0.014}^{+0.015}$\\ 
	$H_{0}$ & $67.97_{-0.70}^{+0.63}$ & $67.97_{-0.68}^{+0.65}$ & $68.08_{-0.70}^{+0.69}$\\
	\hline 
\end{tabular} 
\caption{Dark matter--neutrino coupling; time-independent scattering cross-section.} \label{Tab:MCMCnuconst} 
\end{center}
\end{table}

\bigskip

\begin{table}[h!]
	\begin{center}
\begin{tabular}{|l|c|c|c|c|} 
	\hline 
	Parameter  &  Jeffreys & Linear flat & Logarithmic flat 1\\ \hline 
	$10^{13}\, u^{0}_{\nu}, \; n=2$ & $< 2.463~(95 \%)$ & $< 2.190~(95 \%)$ & $<0.839~(95 \%)$\\ 
	$100~\omega_{b}$ &  $2.219_{-0.017}^{+0.016}$  & $2.220_{-0.016}^{+0.016}$ & $2.224_{-0.016}^{+0.016}$ \\ 
	$\omega_{\textrm{cdm}}$ & $0.1196_{-0.0015}^{+0.0015}$ & $0.1195_{-0.0015}^{+0.0015}$ & $0.1193_{-0.0015}^{+0.0014}$ \\ 
	$100~\theta_{s}$ & $1.042_{-0.00031}^{+0.00031}$ & $1.042_{-0.00032}^{+0.00030}$ & $1.042_{-0.00030}^{+0.00030}$ \\ 
	$\ln\left(10^{10}A_{s}\right)$ & $3.045_{-0.027}^{+0.025}$ & $3.047_{-0.026}^{+0.026}$ & $3.051_{-0.026}^{+0.025}$ \\ 
	$n_{s}$ & $0.9562_{-0.007}^{+0.006}$ & $0.9576_{-0.0063}^{+0.0066}$ & $0.9617_{-0.0053}^{+0.0059}$ \\ 
	$\tau_{\textrm{reio}}$ & $0.05894_{-0.014}^{+0.014}$ & $0.05943_{-0.014}^{+0.014}$ & $0.06043_{-0.014}^{+0.014}$ \\ 
	$H_{0}$ & $67.96_{-0.70}^{+0.64}$ & $67.96_{-0.68}^{+0.65}$ & $68.04_{-0.66}^{+0.67}$ \\ 
	\hline 
\end{tabular} \\ 
\caption{Dark matter--neutrino coupling; scattering cross-section scales as $\propto T^2$.} \label{Tab:MCMCnuT2}
\end{center}
\end{table}



\bibliography{ref}

\bibliographystyle{utcaps}

\end{document}